\documentclass[sigconf]{acmart}
\AtBeginDocument{%
  }

\usepackage[T1]{fontenc}
\usepackage{xcolor}
\usepackage{colortbl}

\copyrightyear{2025}
\acmYear{2025}
\setcopyright{rightsretained}
\acmConference[CHI EA '25]{Extended Abstracts of the CHI Conference on
Human Factors in Computing Systems}{April 26-May 1, 2025}{Yokohama, Japan}
\acmBooktitle{Extended Abstracts of the CHI Conference on Human Factors in
Computing Systems (CHI EA '25), April 26-May 1, 2025, Yokohama,
Japan}\acmDOI{10.1145/3706599.3720084}
\acmISBN{979-8-4007-1395-8/2025/04}

\begin{document}

\title{The EnviroMapper Toolkit: an Input Physicalisation that Captures the Situated Experience of Environmental Comfort in Offices}

\author{Silvia Cazacu}
\authornote{Both authors contributed equally to this research.}
\email{silvia.cazacu-bucica@kuleuven.be}
\orcid{0000-0002-7952-0919}
\affiliation{%
  \institution{KU Leuven}
  \city{Leuven}
  \country{Belgium}}

 \author{Stien Poncelet}
 \authornotemark[1]
\email{stien.poncelet@kuleuven.be}
\orcid{0000-0003-3248-6781}
\affiliation{
  \institution{Centre for Environment and Health - Dept. of Public Health and Primary Care, KU Leuven}
  \city{Leuven}
  \country{Belgium}
}

\author{Emma Feijtraij}
\email{emma.feijtraij@gmail.com}
\orcid{0009-0003-0545-0975}
\affiliation{%
  \institution{KU Leuven}
  \city{Leuven}
  \country{Belgium}}

\author{Andrew Vande Moere}
\email{andrew.vandemoere@kuleuven.be}
\orcid{0000-0002-0085-4941}
\affiliation{
  \institution{Research[x]Design, KU Leuven}
  \city{Leuven}
  \country{Belgium}
}

\begin{abstract}


The environmental comfort in offices is traditionally captured by surveying an entire workforce simultaneously, which yet fails to capture the situatedness of the different personal experiences.  
To address this limitation, we developed the \textit{EnviroMapper Toolkit}, a data physicalisation toolkit that allows individual office workers to record their personal experiences of environmental comfort by mapping the actual moments and locations these occurred. By analysing two in-the-wild studies in existing open-plan office environments (N=14), we demonstrate how this toolkit acts like a situated input visualisation 
that can be interpreted by domain experts who were not present during its construction. 
This study therefore offers four key contributions: (1) the iterative design process of the physicalisation toolkit; (2) its preliminary deployment in two real-world office contexts; (3) the decoding of the resulting artefacts by domain experts;
and (4) design considerations to support future input physicalisation and visualisation constructions that capture and synthesise data from multiple individuals. 


\end{abstract}



\begin{teaserfigure} 
    \centering
    \includegraphics[width=1\linewidth]{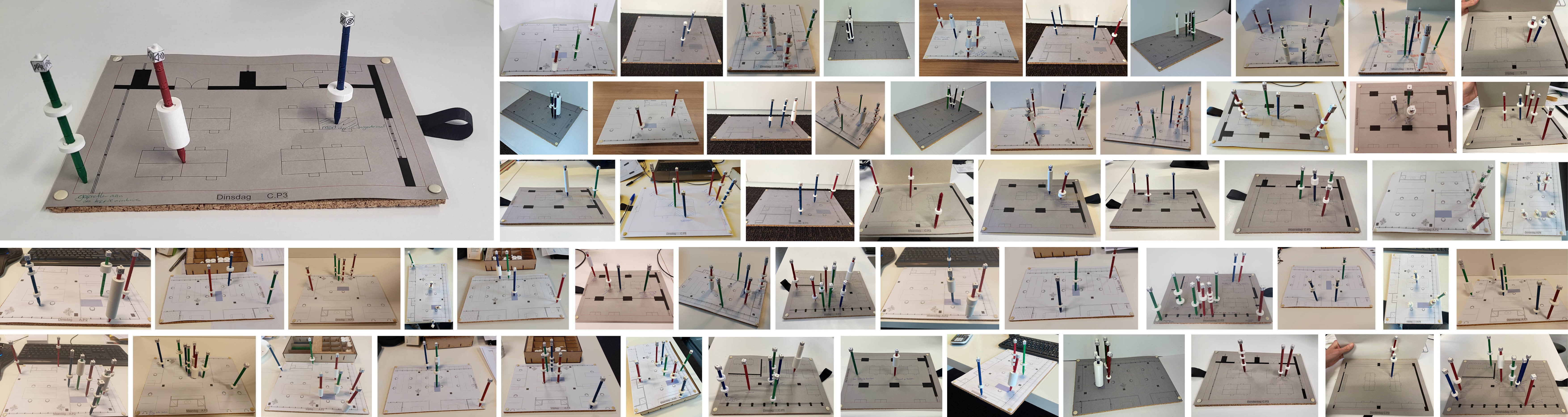}
    \caption{Fourteen office workers constructed 57 input physicalisation artefacts across two open-plan office environments using the EnviroMapper Toolkit, capturing 417 situated experiences of environmental comfort (office A = 305; office B = 112). 
    Negative experiences, marked by red pins (\(43\%\)),
    were mostly related to acoustics (\(56.7\%\)) and temperature (\(17.2\%\)), while positive experiences, marked by green pins (\(28\%\)), were linked to social interaction (\(47.9\%\)) 
    and the presence of nature (\(30.2\%\)). Most proactive changes to the environment, marked by blue pins (\(29\%\)), were related to reducing noise (\(52\%\)). }
    \Description{This image shows an overview of 57 input physicalisation artefacts created by 14 office workers using the EnviroMapper Toolkit in two open-plan office environments. The artefacts, displayed in various tabletop setups, represent 417 situated experiences of environmental comfort, with data focusing on for example acoustics, temperature, social interaction and nature. Each artefact uses coloured pins, rings, cubes and a printed floor plan to visualise insights}
    \label{fig:teaser}
\end{teaserfigure}

\maketitle

\section{Introduction} 

Office workers spend approximately 40 percent of their waking time at work \cite{bodin2019}. Acknowledging this has led to a growing number of \textit{environmental comfort} studies  \cite{Klepeis2001} which aim to understand how office workers experience aspects of the built office environment - such as acoustics, temperature, air quality and lighting. The findings from these studies are utilised by domain experts -  including architects, designers and building managers - to provide evidence-based guidance for future redesigns or modifications to Building Management Systems (BMS) control strategies \cite{Shi2013}. 
These studies typically combine subjective methods, like surveys, with objective data from sensors and monitoring systems to validate or challenge the findings \cite{Lee2023}. However, this approach tends to focus on general insights, 
overlooking the situated aspects (e.g. time and location) 
that dynamically shape personal experiences  \cite{Jayathissa2020}. 
To create more effective, tailored interventions, we need approaches that capture personal situated experiences of environmental comfort, while ensuring their interpretability for domain experts. 
 
We propose input physicalisation - a physical form of `\textit{visual representation that is designed to collect and/or modify new data rather than encode pre-existing datasets}' \cite{inputvis2024} - as a suitable approach to capture situated experiences. 
We presume that the construction of input physicalisation engages office workers in an easily relatable \cite{zhao_embodiment_2008} and enjoyable \cite{hogan2018data} process such as assembling tokens \cite{panagiotidou_co-gnito_2022}, grouping blocks \cite{brombacher_sensorbricks_2024} or manipulating tangible controls \cite{sturdee_exploring_2023}. This process results in a physical artefact \cite{jansen_opportunities_2015} that allows cognitive understanding of data \cite{jansen_evaluating_2013}, making data interpretable as the physical representation engages multiple senses \cite{hogan2012does} and is experienced with the entire human body \cite{hornecker_design_2023}. 
This approach has already been used to collect and modify data on activities \cite{data_badges_2020}, habits \cite{friske_entangling_2020} or preferences \cite{edo_2023}, thus we believe it will allow office workers to create meaning from personal experiences \cite{constructing2014}; to align these meanings 
with others \cite{cairn2017}; and to enrich these meanings with details about the social and physical context where they are situated \cite{bressa2021s}.

We thus introduce the \textit{EnviroMapper}, an input physicalisation toolkit that allows individual office workers to record their personal experiences of environmental comfort by mapping
the actual moments and locations these occurred. In two preliminary in-the-wild studies, 14 office workers and 5 domain experts used the toolkit in the context of two open-plan office environments. Through a thematic analysis of interviews and photos of the 57 physical artefacts (see Figure \ref{fig:teaser}), 
we demonstrate that the toolkit effectively captured detailed environmental comfort experiences, however challenges arose in encoding personal experiences and managing social sensitivities in shared office spaces. Additionally, the intricacy of the
resulting physical artefacts increased cognitive load during decoding, underscoring the need for a balance between data richness and ease of interpretation. 
This study offers four key contributions: (1) the iterative design process of the physicalisation toolkit; (2) their preliminary deployment in two real-world office contexts; (3) the decoding of the resulting artefacts by domain experts; and (4) a set of design considerations meant to support future input physicalisation and visualisation constructions that capture and synthesise data from multiple individuals. 

\section{Related Work} 

Surveys are a standard and widely accepted method for evaluating environmental comfort \cite{sakellaris2016,al2016,carrer2018}. Traditional paper-based surveys \cite{Csikszentmihalyi1977,Muller2014} capture detailed snapshots, but lack contextual richness. Annual surveys \cite{Boubekri1993,Lee2021} focus on topics like wellbeing \cite{Koch2012}, yet their infrequent nature limits their capacity to capture in-the-moment feedback \cite{deWaal2014}. Mobile surveys \cite{VanBerkel2017} address this by prompting individuals in the moment, which reduces memory bias but raises concerns about notification overload and privacy risks \cite{Gouveia2013, Muller2009, Raento2009}. Moreover, surveys are often augmented with sensor technologies that objectively measure and monitor environmental aspects like temperature, lighting and sound. Examples include Office Agents \cite{Stamhuis2021} and Break-Time Barometer \cite{Kirkham2013}, which combine environmental data with individual behaviour. Mobile sensors like SensorBadge \cite{Brombacher2022} provide personalised insights, but often lack effective localisation. Hybrid systems \cite{Huang2014,Salamone2021} address these gaps, however they can be cost-prohibitive and overlook subjective experiences.

Input visualisation systems \cite{postervote2014,sens-us2015,marketmodel2014,myposition2014,viewpoint2012} allow for localised data collection without explicit instructions \cite{smalltalk}. They collect subjective data from individuals and link personal input to crowd-level data \cite{VoxBox}, though the immediate visibility of these data raises privacy concerns \cite{Pub_Vis2015}, which can be avoided by hiding data during input \cite{inputvis2024}. In their physical form, they reduce cognitive overload through physical metaphors, however they lack the precision of digital visualisations \cite{click2024} while short time windows for data input limit the effectiveness of their analysis later on \cite{roam-io2019}. When used in office environments, such input data collection systems foster social connectedness but may pressure individuals to conform to social norms \cite{mood_squeezer2015}, potentially altering their behaviour when their co-workers observe their responses \cite{edo_2023}.
Considering these aspects, we identify opportunities to address the gaps in traditional surveys by using input physicalisation to capture real-time, in-the-moment experiences by reducing memory bias and collecting rich contextual data, while being mindful of privacy concerns and potential biases from social interactions.

\section{Design}

\subsection{Conceptual design.}
The conceptual design of the EnviroMapper Toolkit is based on the Environmental Demands-Resources (ED-R) framework \cite{Roskams2021, Roskams2020}, 
which describes how office workers experience environmental comfort  
through \textit{environmental demands, resources} and \textit{crafting}.
Environmental demands represent aspects in the built office environment that cause physiological and/or psychological strain, making it challenging 
to achieve 
goals. Examples include poor \textit{air quality}, extreme \textit{temperature}, inadequate \textit{lighting}, disruptive \textit{acoustics}, bad \textit{ergonomics} and lack of visual or acoustic \textit{privacy} \cite{Roskams2021}. Conversely, environmental resources stimulate engagement, enhance motivation and help office workers to recover from stress. Examples include the presence of \textit{nature}, the ability to \textit{personalise}, appealing \textit{aesthetics} and spaces that foster \textit{social interactions} \cite{Roskams2021}. When office workers make
(pro-)active changes to either reduce environmental demands (e.g. wearing headphones to block distracting noise) or enhance environmental resources (e.g. opening curtains to enjoy a view of the outdoors), they engage in environmental crafting \cite{Roskams2021b}. 

\subsection{Iterative Design Process.}
\textit{\textbf{Low-Fidelity Prototyping.}}
We first translated the ED-R framework into the following data dimensions: type of experience (environmental demand or resource), environmental aspect (e.g. acoustics, nature, temperature), location, intensity, duration and time of day. To understand how these six data dimensions could be best combined into an input physicalisation toolkit that supports intuitive, fast, and repeated data input across multiple days, either individually or in groups, we developed three low-fidelity prototypes (shown in Appendix Figure \ref{fig:low-fi}), which we evaluated internally for (re-)usability, transportability and interpretability. The first prototype consisted of a timetable with stickers representing each data dimension. Although this was intuitive for users, it did not indicate the location of the experiences mapped. Thus, we developed the second prototype: an abstract representation of space where coloured sticks could be placed into a soft base. Even though the sticks indicated the location of the experiences mapped, this could only be subjectively interpreted by the people who mapped it. The third prototype therefore consisted of a detailed floor-plan where sticks connected experiences to a specific location, while discs added a time dimension. Even though the time duration and specific moment of the day was still difficult to interpret, we decided to continue developing this prototype into a mid-fidelity one. This prototype allowed workers to individually map their experiences at their own desk, to move the resulting artefact to different locations in case of flexible desk arrangements, and to easily clear the toolkit at the end of each day.

\begin{figure*} 
    \centering
    \includegraphics[width=1\linewidth]{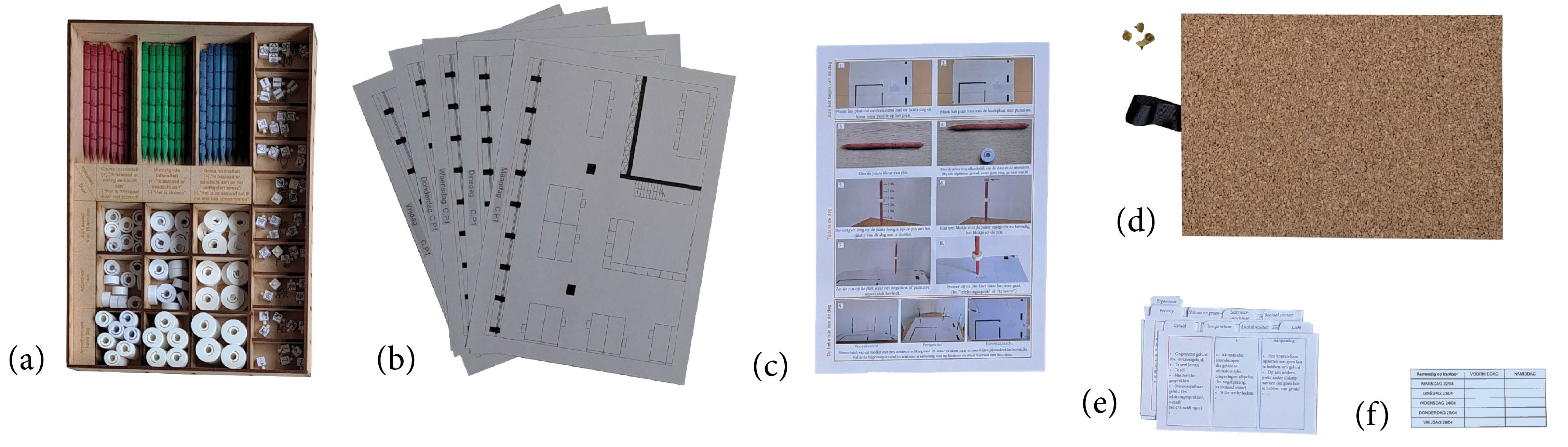}
    \caption{The high-fidelity prototype of the EnviroMapper Toolkit featured: (a) a components box containing red, blue and green pins engraved with time markers, blocks to specify the type of environmental aspect represented and rings of varying diameter and height to indicate intensity and duration; (b) printed floor plans; (c) an instruction sheet; (d) the box cover consisting of a cork board to secure the plan with thumb tacks; (e) a booklet with examples of situated experiences per environmental aspect; and (f) a printed attendance table for tracking participation.}
    \Description{This image shows the high-fidelity prototype of the EnviroMapper Toolkit, divided into labelled components. (a) A components box containing red, blue and green pins engraved with time markers, blocks for identifying environmental aspects and rings of different sizes to indicate intensity and duration. (b) Printed floor plans for mapping. (c) An instruction sheet with step-by-step visual guidance. (d) A cork board serving as a box cover for securing plans with thumb tacks. (e) A booklet with examples of situated experiences for various environmental aspects. (f) A printed attendance table for tracking participation.}
    \label{fig:finaldesign}
\end{figure*}

\textit{\textbf{Mid-Fidelity Prototyping.}}
We created a mid-fidelity prototype using a foam board with a furnished floor plan (see Appendix Figure \ref{fig:mid-fi}). Red and green pins marked environmental demands and resources, rings of varying diameters indicated duration and stickers specified the environmental aspect. Initial tests with five friends and family members 
led to several adjustments: scaling up the floor plan for more precise mapping, improving ring sliding and securing stickers with thumb tacks, adding space for notes and sketches, and introducing thicker rings for day-long durations. Moreover, since the flowchart - illustrating the construction process - was too complex and seldom used, we replaced it with an illustrated instruction sheet and a set of cards with examples of situated experiences per environmental aspect. Furthermore, blue pins were added to represent environmental crafting as adaptive behaviours were observed. Finally, recognising that some environmental demands could be perceived positively (e.g. a fresh breeze), we continued to use green pins for positive experiences and red pins for negative experiences. 
\textit{\textbf{High-Fidelity Prototyping}.}
We developed six copies of a high-fidelity prototype using laser cut, printed and 3D-printed components, as presented in Figure \ref{fig:finaldesign}. The 3D-printed pins, rings and blocks improved usability and allowed for precise representation of time, intensity and duration. Several iterations of the 3D-printed components are presented in Appendix Figure \ref{fig:high-fi}. The laser-cut box provided organised compartments for easy retrieval of the components, while printed floor plans could be replaced daily for repeated use. 

\subsection{Mapping Situated Experiences with the EnviroMapper.}
Office workers mapped their experiences by following predetermined encoding rules to construct a physical artefact throughout the day. The encoding rules were printed on the instructions sheet (see Appendix, Figure \ref{fig:encoding-rules}) and a set of cards that explained the different types of environmental demands, resources and crafting experiences. According to the instructions sheet, at the beginning of each day, office workers attached the floor plan of their immediate office environment to a cork board and marked their location on the floor plan with a cross. Next, when a situated experience occurred, they chose colour-coded pins - red for negative experiences, green for positive experiences and blue for crafted experiences. Then, they chose a ring according to the duration and intensity. Next, they placed the ring on the pin according to the precise time of the day when the experience occurred. Once the pin and ring were assembled, they were placed onto the board at the precise location where the experience occurred. Additional blocks, indicating the type of environmental aspect experienced were then placed on top of the pin. At the end of the day, office workers documented their resulting physical artefact by taking photographs from three angles: front view, front perspective and top view. Finally, they cleared the toolkit for the next day by returning all pins, rings and blocks to the box.

\section{Methodology}

\textbf{\textit{In-the-Wild Studies with Office Workers.}}
The two in-the-wild studies conducted in office A and office B (pictures and plans included in Appendix Figure \ref{fig:office-plans}) started with an introduction session where office workers were introduced to the study and toolkit through a demonstration, signed informed consents and completed an online demographic questionnaire. Office workers then used the toolkit for one week while sending photographs of their resulting artefacts to the researchers daily. Afterwards, each office worker took part in an one-hour semi-structured interview to discuss their experiences, toolkit use, interactions with others and overall user-friendliness. Prior to the interviews, researchers reviewed the artefact pictures to identify insights and unusual usage patterns.

\textbf{\textit{Data Interpretation Sessions with Domain Experts.}}
Five domain experts with background and expertise in architectural design interpreted the resulting physical artefacts of offices A and B during two sessions (detailed in Appendix Figure \ref{fig:data-interpretation}. Session 1, lasting one hour, involved two domain experts unfamiliar with office A. This session included an introduction of the toolkit and office A; an analysis of artefact pictures of office A supported by interview quotes and a replicated physical artefact; and a discussion about the meaning of the data captured and the interpretability of the physical artefacts generated with the toolkit. To interpret the data captured, the experts performed the following tasks: 1) \textbf{identify} pin colour, placement, ring placement on pin and blocks; 2) \textbf{compare} pin colours, pin placement, ring diameter and height; 3) \textbf{count} pin colours, ring diameters, heights, placement and blocks. As they completed these tasks, they made annotations on a participant sheet that depicted all physical artefacts constructed by that participant (find an example in Appendix Figure \ref{fig:expert-annotations}).
Session 2 consisted of three interviews with domain experts with varying knowledge of the toolkit and office B. Like Session 1, this session followed the same phases and tasks, while adding the task of developing design proposals aimed at improving the environmental comfort of office B, based on the data interpreted from the physical artefacts.

\textbf{\textit{Data Analysis.}}
We analysed two types of data: (1) pictures of the physical artefacts of office A and office B; and (2) transcriptions of audio-recordings from the individual interviews with office workers and the two sessions with domain experts. The pictures of the physical artefacts were analysed quantitatively to complement the qualitative data originating from the interviews. These interviews were analysed by the third author using thematic analysis \cite{Braun2006} to identify, analyse and report patterns, resulting in six sub-themes and two themes.


\begin{figure*} [h]
    \centering
    \includegraphics[width=1\linewidth]{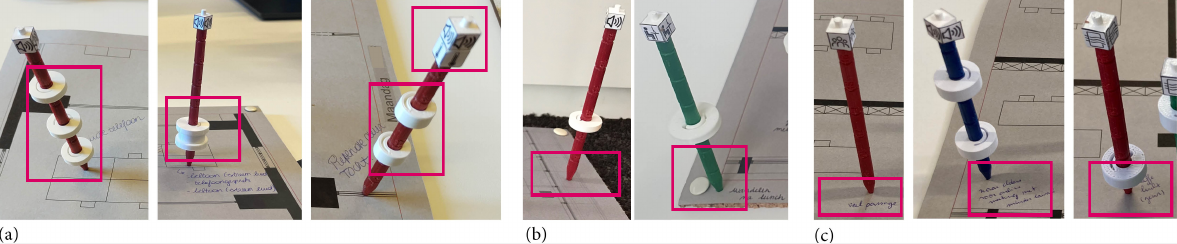}
    \caption{Appropriating the EnviroMapper by adapting the encoding rules: (a) multiple rings representing repeated events, varying intensities or combined environmental aspects; (b) pins placed on the border of the plan to signify environmental aspects outside the depicted area; and (c) diverse interpretations of environmental aspects (e.g. social interaction for crowded office, acoustics for moving to another space for meeting and air quality for musty smell).}
    \Description{A series of photos showcasing the appropriation of the EnviroMapper tool through customised encoding rules. (a) Pins with multiple rings symbolise repeated events, varied intensities, or combined environmental aspects, marked on a floor plan. (b) Pins placed at the edges of the plan represent environmental aspects occurring outside the mapped area. (c) Pins illustrate diverse interpretations of environmental factors such as social interaction in crowded spaces, acoustics when relocating for meetings, and air quality issues like a musty smell. Pink rectangles highlight key areas in each image.}
    \label{fig:appropriation}
\end{figure*}

\section{Results}
We recruited 14 office workers for the in-the-wild studies (8 for office A; 6 for office B), consisting of 9 men and 5 women, aged 24 to 57 (mean = 34.79, SD = 11.29), that used the toolkit for one week in their office environment, by constructing one new physical artefact each day. Furthermore, 5 domain experts evaluated the photos depicting the physical artefacts. In this section, each office worker and domain expert is identified by a code (e.g. OA1) that combines their role (O = office worker; D = domain expert) and the office where the toolkit was deployed (A = office A; B = office B).
%

\subsection{Appropriation Strategies of the EnviroMapper Toolkit.}
\label{sec:mapping_approaches}

When asked about a digital alternative to map situated experiences, most office workers in office B (\(83\%\)) answered that they would prefer digital over physical mapping when it comes to a faster use and fewer privacy concerns. However, office workers highlighted that the added value of the physical representation compared to a digital once consisted in its ability to provide deep insights about recurring experiences. From this perspective, the toolkit was described as more engaging in use; more visually appealing for presenting results; and more effective in eliciting reflections on the built environment compared to a survey.
Overall, the toolkit was rated highly for its clarity by the office workers, however office workers expressed concerns about not knowing how to map specific experiences that did not fit the six data dimensions. Consequently, they appropriated the toolkit by developing unique strategies to overcome these challenges.

\textbf{\textit{Appropriation by Adapting the Encoding Rules.}} \label{encoding rules}
Most office workers (8/14) struggled to provide context that could not be captured with the toolkit, which they overcame by adapting the encoding rules (see Figure \ref{fig:appropriation}). For example, they stacked multiple rings on a pin to represent repeated instances of the same environmental aspect rather than separate experiences. They also combined rings of varying intensities to map details, such as a loud ringtone followed by a phone call, or stacked cubes to capture co-occurring environmental aspects (e.g. a noise and a breeze when a door opened). Additionally, the majority (8/14) wrote notes near pins to aid recall, clarify experiences or add context for later interpretation.

\textbf{\textit{Appropriation by Delaying Mapping.}}
Several office workers (3/14) 
opted to make mental notes of their experiences and mapped them onto the toolkit later during idle moments at work, on short breaks or even at fixed times they planned in advance: \textit{`I am rather a planning person. I kept in mind that I had to be aware of what was disturbing me or where it came from, and I also kept track of the things in my mind, for each half day, I mapped the experiences of what happened'} (OA7). 
Other office workers (3/14) 
made notes on their computers as they experienced something to then transfer these notes onto the toolkit at the end of the work day. Delayed mapping was preferred for being less distracting while working, more private, saving desk space and increasing assembling efficiency.

\textbf{\textit{Appropriation through Self-Reflection.}}
Most office workers (8/14) indicated that the toolkit increased their awareness of different environmental aspects: `\textit{There are a few aspects of the environment that I appreciate more now, like the positive aspects where I did not think about before}' (OB2). 
Six office workers 
indicated that they used the toolkit to reflect on their environment and daily disturbances: `\textit{Tuesday, when I was sitting in a different space, I already knew that I don’t like sitting there, but now I could really think about why I don't like it. I also work in other offices sometimes, so now I understand more why I like being at one office or another}' (OB6).

\textbf{\textit{Appropriation as Social Adaptation.}}
While most office workers primarily discussed the toolkit with co-workers who were also using it, such as addressing `\textit{what is or is not considered as a disturbing factor to indicate on the map}' (OA6), one noted that that these discussions led them to map previously overlooked environmental aspects.
Several office workers (6/14) 
noted that co-workers who were not using the toolkit, once familiar with it, were able to interpret experiences and even showed interest in identifying disturbances: `\textit{They joked a bit that they had to be well-behaved around me or I would put a red pin on their place}' (OA5). Two office workers 
indicated that these interactions encouraged them to engage more actively with the toolkit.
However, five office workers 
felt embarrassed while mapping experiences involving specific co-workers. To avoid this, they used the toolkit differently by placing blue pins instead of red ones for negative experiences or adding notes and pins in the evening when others were not around.

\begin{figure*} [ht]
    \centering
    \includegraphics[width=1\linewidth]{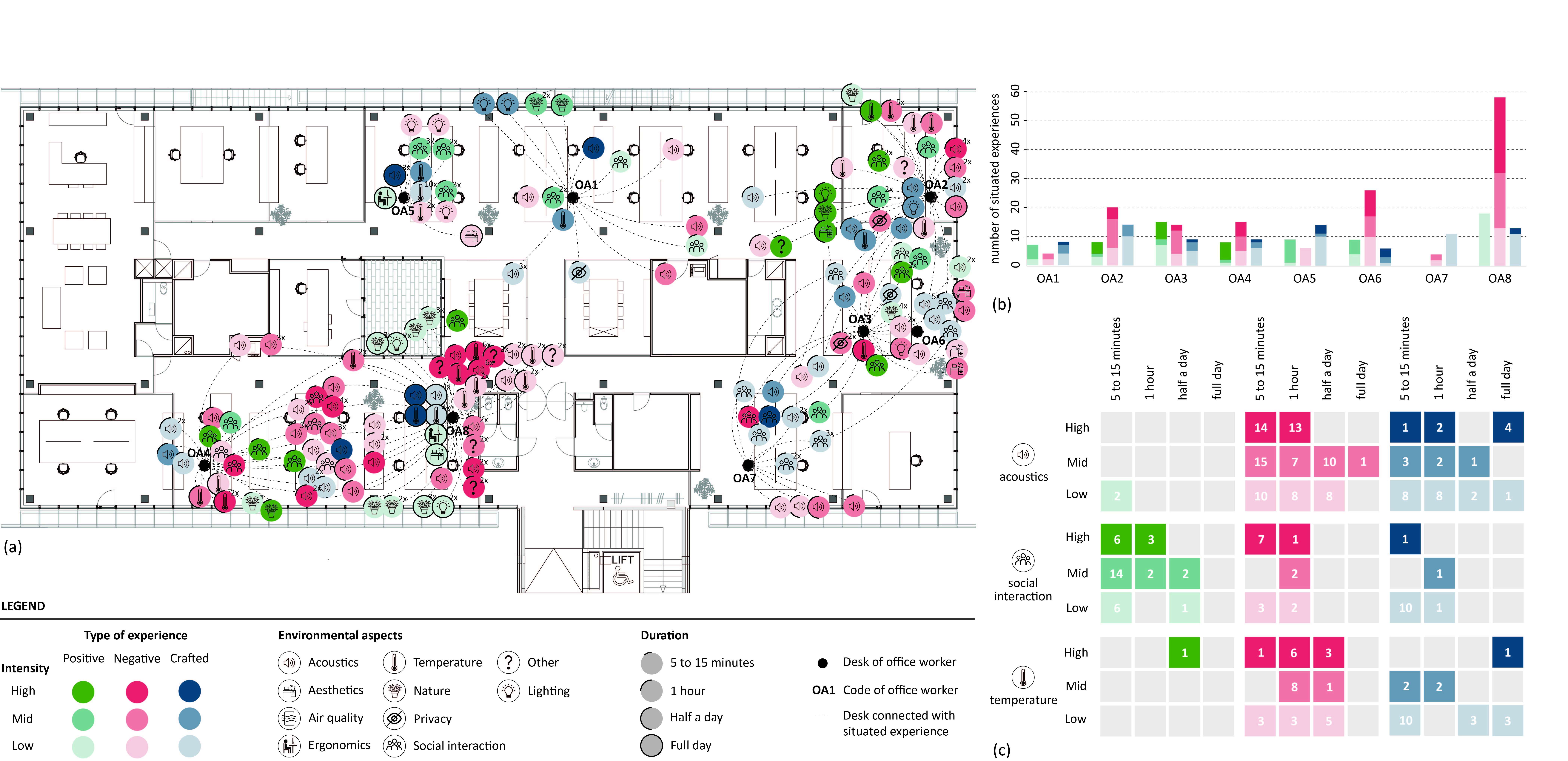}
    \caption{By visually representing the situated experiences mapped by 8 office workers from office A, we are able to synthesise that: (a) acoustic distractions mainly originate around the desks of other co-workers, while positive interactions are experienced at one's own desk; (b) the individual approaches and frequencies for mapping experience and intensity reveal insights into office workers' personalities; (c) the most frequently mapped environmental aspects are acoustics, social interaction and temperature, and they are mapped in relation to their duration, intensity and type. A similar synthesis for office B appears in Appendix Figure \ref{fig:heatmap-appendix}.}
    \Description{A detailed visualisation of mapped situated experiences in the office environment of office A, paired with analytical charts. The office floor plan (a) features a heatmap with colour-coded markers for office worker experiences linked to environmental aspects like acoustics, social interaction, and temperature, categorised by intensity (high, mid, low) and type (positive, negative, crafted). A bar chart (b) highlights experience types and intensities per office worker, while a grid chart (c) summarises key environmental aspects by duration and intensity. A legend decodes the symbols used for environmental aspects, intensity levels, and experience duration.}
    \label{fig:heatmap}
\end{figure*} 

\subsection{Interpretation of the EnviroMapper Toolkit.} \label{sec:interpretability}
Overall, domain experts pointed out that analysing the artefacts `\textit{is not necessarily more work than, say, an interview with each person individually}' (DB3), because the toolkit immediately represents data once it is collected. In comparison, domain experts explained that using a survey would be `\textit{more difficult to make conclusions than from the pictures of the used toolkits'} (DB2). However, they also expressed difficulties with interpreting the data collected due to (1) digital reproduction of physical artefacts; (2) diverse artefact embodiments; and (3) diversity of data dimensions. 

\textbf{\textit{Analysing Images of the Physical Artefacts.}}
For some domain experts (3/5), having the physical toolkit at hand during analysis was helpful for examining details, as they often misinterpreted pin colours due to insufficient colour saturation and thin pin shapes, which was also reported by several office workers.
Three domain experts 
mentioned that the pictures used for analysis were sometimes unclear or too small to distinguish environmental aspects effectively. Additionally, while perspective and front-view pictures were primarily utilised, top-view pictures provided little added value because the environmental aspects were then not visible anymore, while the white rings were difficult to distinguish from the white background of the office plan.

\textbf{\textit{Discrepancy in Visual Variables.}}
Several domain experts (3/5) highlighted challenges in interpreting intensity differences because the rings representing small intensity and long duration were disproportionately larger in surface area compared to those indicating large intensity and short duration. This discrepancy made long-duration rings more visually prominent, potentially overstating their impact compared to shorter, high-intensity experiences. To overcome this limitation, two experts emphasised the importance of follow-up and note-taking to explain the placement of pins.

\section{Discussion}


As Figure \ref{fig:heatmap} shows, the EnviroMapper Toolkit captured detailed situated experiences of environmental comfort in relation to their locations, while revealing diverse mapping frequencies and approaches. Yet, encoding the meaning of personal experiences into data is challenging. Furthermore, the process of constructing data physicalisation is perceived as an experiential act in and by itself. These aspects combined reveal the strengths and shortcomings of the toolkit.


\textit{\textbf{Tensions between Predetermined Encoding Rules and Personal Experiences of Environmental Comfort.
}}
Our results indicate that the encoding rules did not allow office workers to capture serendipitous experiences of human comfort.
As office workers delayed mapping and recombined pins and rings (see Section \ref{encoding rules}), they changed the order of encoding the six data dimensions and even skipped several encoding rules to capture different experiences simultaneously. This challenge was exacerbated when office workers employed varying data dimensions to encode the exact same experiences, as shown in Figure \ref{fig:appropriation}. This corresponds to similar findings that emphasised by how people use different encoding rules to represent similar personal datasets because personal experiences are inherently ambiguous \cite{self-reflection2018}.
The office workers tendency to override encoding rules was caused by the wish to control the encoding with strict rules, so that the resulting artefacts could be decoded by others. It has been shown that domain experts struggle to decode data encoded through unfamiliar rules (see Section \ref{sec:interpretability}) \cite{panagiotidou_co-gnito_2022}. However, this limited the ways that the toolkit could be appropriated.
One way to overcome these boundaries was to take notes for delayed mapping, as notes can be considered part of the physicalisation itself \cite{walny2015}.
At the same time, qualitative data, such as explanations, cannot be visually encoded without some synthesis.
Overall, a balance must be found between using intuitive encoding rules that are easy to follow and decode, allowing an artefact to be internally and externally consistent, while capturing the richness of qualitative experiences, which vary among individuals and allow for more aspects than what data can capture.





\textit{\textbf{Tensions between Artefact Construction and the Shared Office Environment.}} Our findings demonstrate that the construction of data physicalisation artefacts in the same office as others biased how office workers captured aspects of environmental comfort that are socially sensitive.
Most office workers felt embarrassed to construct artefacts that physically represent experiences related to other co-workers (e.g. a phone call or a passer-by) because they were afraid that the artefacts could be interpreted by those co-workers (see Section \ref{encoding rules}). This provides additional evidence that office workers do not want to violate social norms in office environments by collecting data that might hurt the feelings of others \cite{mood_squeezer2015}.
To mitigate this discomfort, office workers adapted their construction process to manage these experiences by swapping red pins with a milder option of a blue pin or by delaying mapping until other workers left.
This can be perhaps explained by how people evaluate and modify their actions to avoid causing conflict or distress to others in social contexts \cite{Coyne1991,Keltner1997}. 
To minimise uncomfortable situations, physicalisations that capture social environments should avoid placing the artefact near highly visible areas like desks or afford delayed data encoding. 

\textit{\textbf{Tensions between Encoding and Decoding.}}
As captured by Figure \ref{fig:heatmap}, our results suggest that while the physical artefact expressed rich personal experiences of environmental comfort, its intricacy imposed cognitive load during decoding. 
The process of meticulously encoding experiences across diverse data dimensions with many ordinal steps and categorical types  changed perceptions of office worker's environment (see Section \ref{encoding rules}). This brings evidence that physicalisation construction fosters reflection by externalising cognitive processes into physical artefacts \cite{self-reflection2018} that are incrementally constructed \cite{constructing2014}. 
However, once encoded, domain experts faced challenges in decoding the artefacts (see Section \ref{sec:interpretability}). We recognised three decoding challenges that can be explained by difficulties in visual perception.
First, the toolkit adhered to recommended channel rankings (starting with position, followed by colour and area \cite{mccoleman2021rethinking_channels}) in principle, but overlooked how the colour was not readily perceptible due to the narrowness of the pins while the icons were not readily perceptible. As a result, domain experts often skipped pin colour and prioritised pin position, followed by ring height and diameter.
Second, the rings perhaps further compounded errors, as they were potentially misinterpreted by volume rather than surface area. It has been shown that such errors occur because people have perceptual difficulties with analysing volume differences \cite{jansen2015psychophysical}. 
Third, the white rings also blended with the white background once the physical artefacts were transformed into images, forcing experts to switch between multiple views of the physical artefact for accurate interpretation of all the data dimensions it embodied. These issues combined require repetitive manual decoding steps that significantly increase cognitive load \cite{ma2019decoding}.
To improve decoding, we should align more closely with the visual cognition principles in visualisation \cite{mccoleman2021rethinking_channels}, perhaps compromising other aspects of the physicalisation artefact (e.g. increasing the pin diameter could reduce the accuracy of the mapped location). Finally, we should  consider digital means of decoding, e.g. analysing images or videos of the artefact via computer vision.





\section{Conclusion}
We developed the EnviroMapper Toolkit that empowers individual office workers to record their personal experiences of environmental comfort by mapping
the actual moments and locations these occurred. Two in-the-wild studies demonstrated that while the toolkit effectively captured detailed, location-based experiences, office workers faced challenges in encoding serendipitous experiences and managing social sensitivities. Additionally, the intricacy of the resulting physical artefacts increased the cognitive load of domain experts during decoding, highlighting the need for a balance between data richness and interpretability.

\begin{acks}

The research of Silvia Cazacu has received funding from the European Union’s Horizon 2020 research and innovation programme under the Marie Sklodowska-Curie grant agreement No. 955569. The opinions expressed in this document reflect only the authors' views and in no way reflect the European Commission’s opinions. The European Commission is not responsible for any use that may be made of the information it contains.

The research of Stien Poncelet is supported by the KU Leuven ID-N project IDN/22/003 `Adaptive Architecture: the Robotic Orchestration of a Healthy Workplace'.

We kindly thank Lybover and Sweco for allowing us to conduct this study at their offices and for their support in managing the practicalities of the study deployment. Finally, we thank all participants and experts for taking the time from their busy schedules to attend and contribute to our research.

\end{acks}

\bibliographystyle{ACM-Reference-Format}
\bibliography{EnviroMapper}

\appendix
\section{Appendix}

\begin{figure} [b]
    \centering
    \includegraphics[width=1\linewidth]{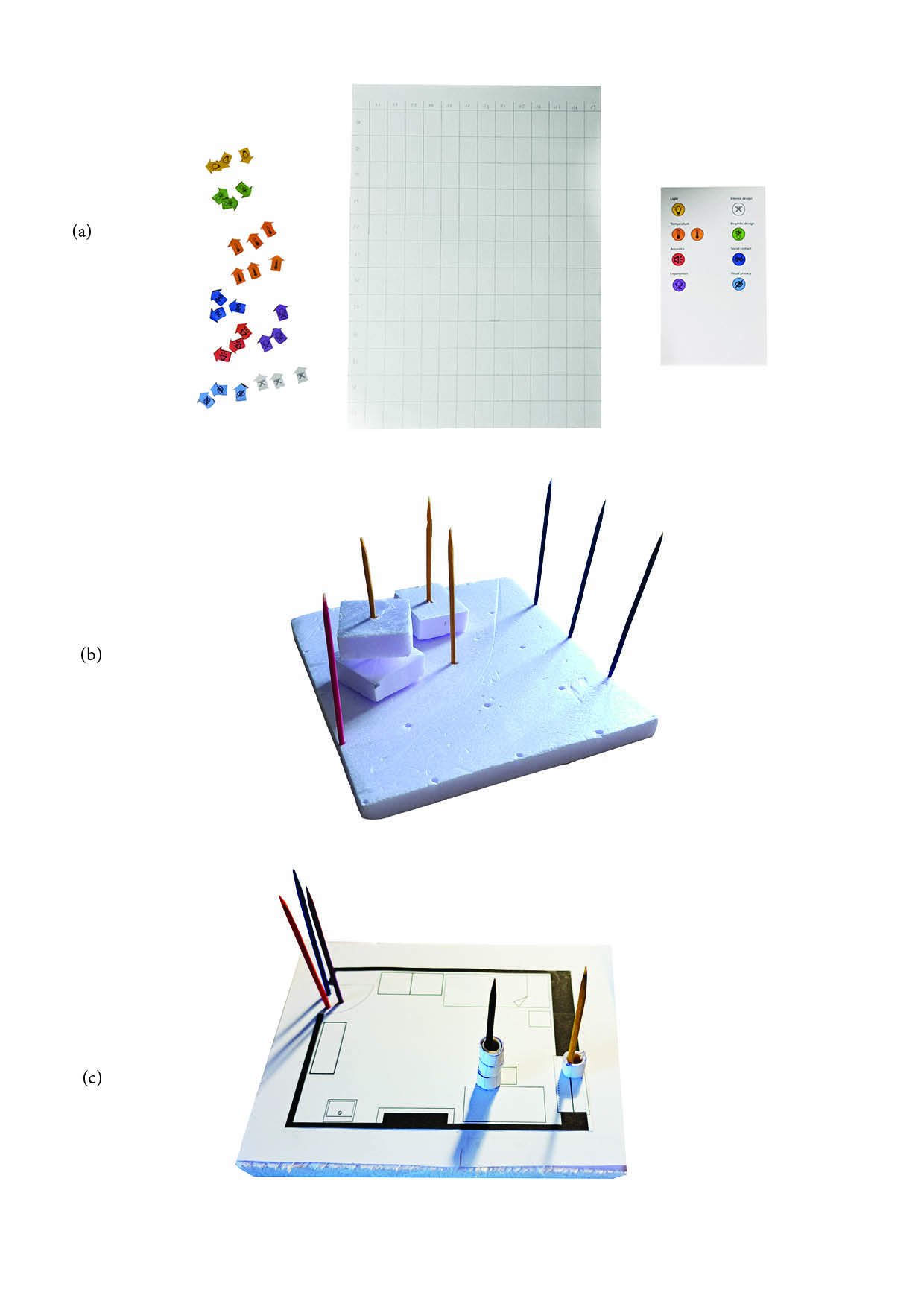}
    \caption{We developed and tested low-fidelity prototypes: (a) timetable prototype featuring a grid sheet, colourful arrows and a legend categorising environmental aspects by symbols and colours; (b) abstract representation prototype utilizing a 3D foam model with upright sticks of various colours and heights to represent different environmental aspects; (c) detailed plan prototype consisting of a floor plan with pins placed at specific locations to indicate environmental aspects.}
    \Description{A comparison of three low-fidelity prototypes for mapping environmental aspects. (a) A time-table-based prototype featuring a grid sheet, colourful arrows and a legend categorising environmental aspects by symbols and colours. (b) A model-based prototype utilizing a 3D foam model with upright sticks of various colours and heights to represent different environmental factors. (c) A plan-based prototype showing a floor plan with pins placed at specific locations to indicate environmental aspects. }
    \label{fig:low-fi}
\end{figure}


\begin{figure*}[htbp]
    \centering
    \includegraphics[width=1\linewidth]{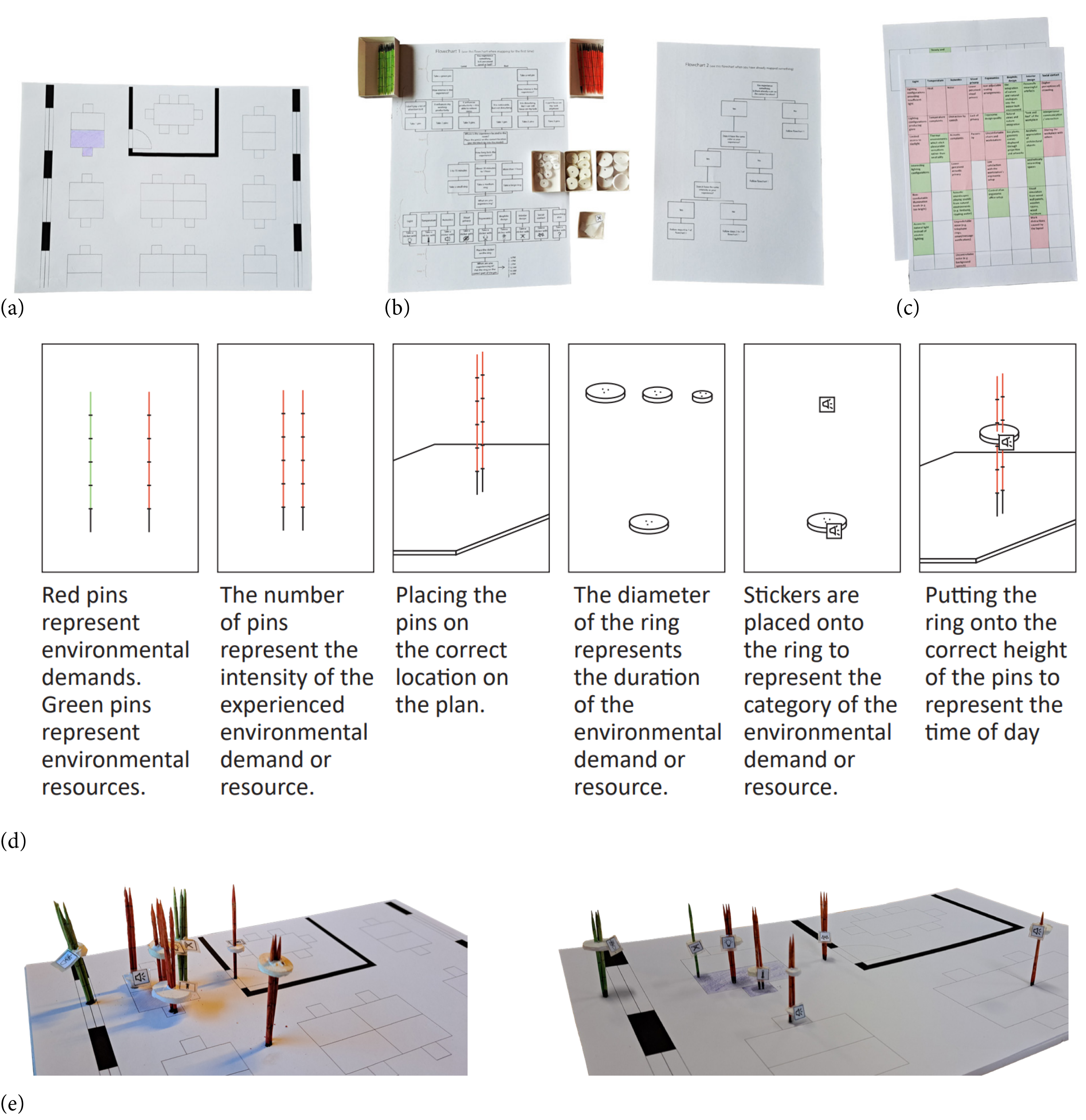}
    \caption{The mid-fidelity prototype did not account for environmental crafting or experience intensity at this point, therefore it relied on the following components: (a) printed office plan on a foamboard; (b) flow chart and 3D-printed components to represent a situated experience: red and green pins (no blue pins) to indicate environmental demands and resources as well as rings to represent the duration; and (c) example of several situated experiences. (d) A flow chart of the encoding rules was added to help users navigate the physicalisation construction process. (e) The physical artefact constructed during an informal evaluation by a family member in an open-plan office environment.}
    \Description{A composite image showcasing a mid-fidelity prototype for mapping situated experiences in an open-plan office. Panel (a) displays a foamboard plan, (b) highlights flow charts, red and green pins for environmental demands and resources and rings for duration representation, and (c) provides an example of situated experiences. Panel (d) illustrates the construction process, including steps for pin placement, ring sizing and categorisation. Panel (e) features physical artefacts with coloured pins, rings and stickers placed on a foamboard during an evaluation with a family member.}
    \label{fig:mid-fi}
\end{figure*}



\begin{figure*}[htbp]
    \centering
    \includegraphics[width=0.75\linewidth]{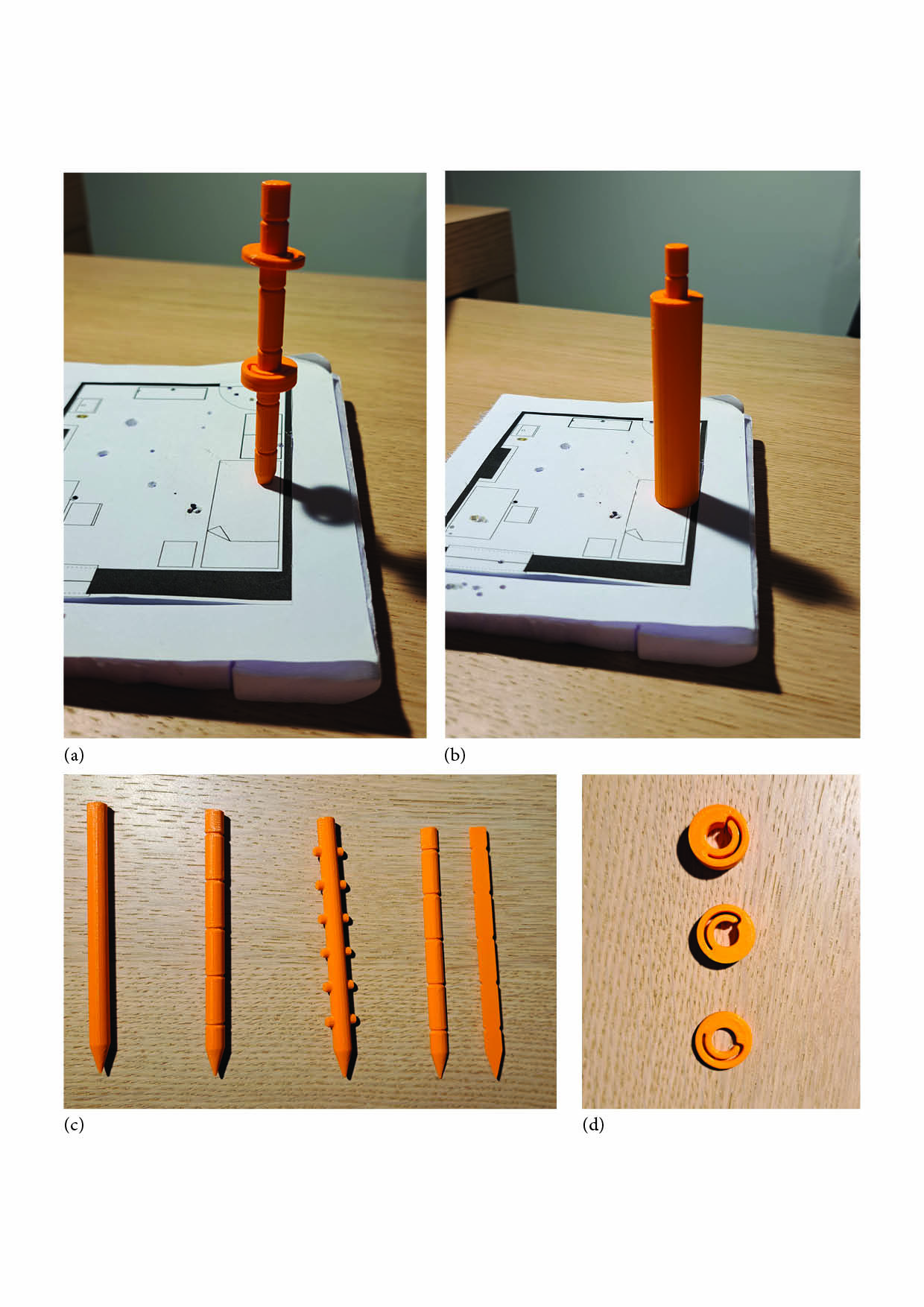}
    \caption{We fabricated and tested several combinations of pins and rings to ensure that the rings could easily fit the pins while keeping a correct placement on the time of the day marker of the pin. (a) An informal evaluation where two experiences of different durations and intensities and (b) one experience of a long duration occur on the same location. (c) We tested different time of day markers on a pin and (d) several ring heights to indicate variations in duration.}
    \label{fig:high-fi}
    \Description{This is a composite figure of four images: images a) and b) show pins with rings of multiple heights and diameters placed on a floor plan to indicate a situated experience. Image c) shows five pins of different sizes and with a different horizontal marks, from indentations to ribs. Image d) shows three rings of different heights and diameters.   }
\end{figure*}



\begin{figure*}[htbp]
    \centering
    \includegraphics[width=0.75\linewidth]{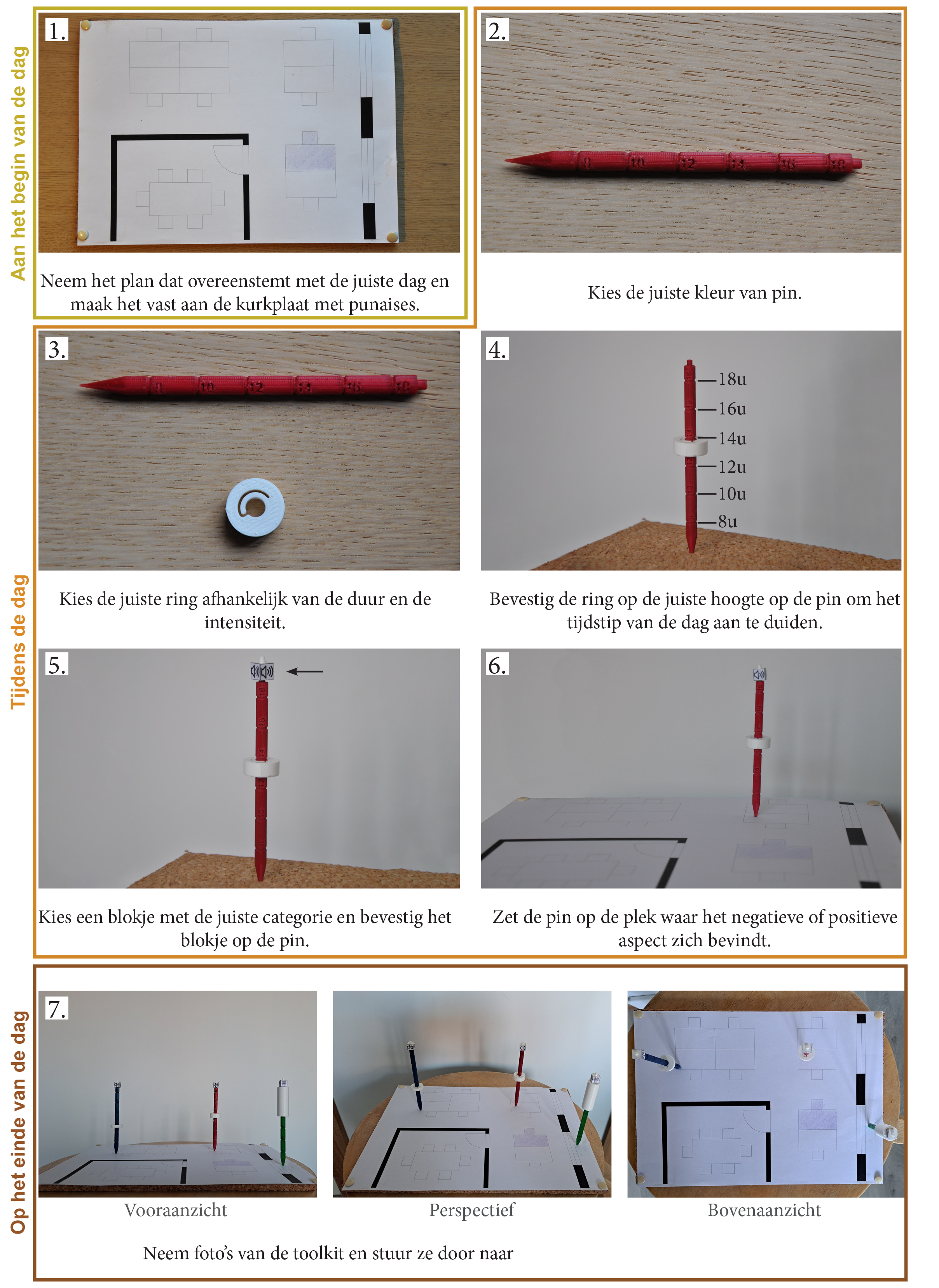}
    \caption{These encoding rules were printed on an instructions sheet and enclosed in the toolkit box. 1) At the beginning of each day, a new printed floor plan should be placed on the cork board. 2) To map a situated experience, first choose the correct pin colour, then 3) choose an appropriate ring height and diameter according to the duration and intensity of the experience and 4) place the ring on the correct time of day marker on the pin. 5) Choose a block indicating the type of environmental demand, resource or crafting experience and place it on top of the pin. 6) Place the pin on the right location on the floorplan. 7) At the end of the day, take 3 photos to document the artefact resulted, from front view, front perspective and top view. Send these photos to us by email.}
    \label{fig:encoding-rules}
    \Description{This is a composite figure of seven images. 1) An empty floor plan. 2) A red pin on a flat surface. 3) A red pin next to a white ring. 4) A red pin with a white ring inserted in a soft surface, with annotated markers representing the time of day. 5) A block added on top of the previous red pin. 6) A red pin with a white ring and a block attached is placed onto the floor plan surface. 7) Three images show a floor plan with a red, a green and a blue pin placed onto its surface, captured from three angles: front view, front perspective and top view.    }
\end{figure*}



\begin{figure*}[htbp]
    \centering
    \includegraphics[width=1\linewidth]{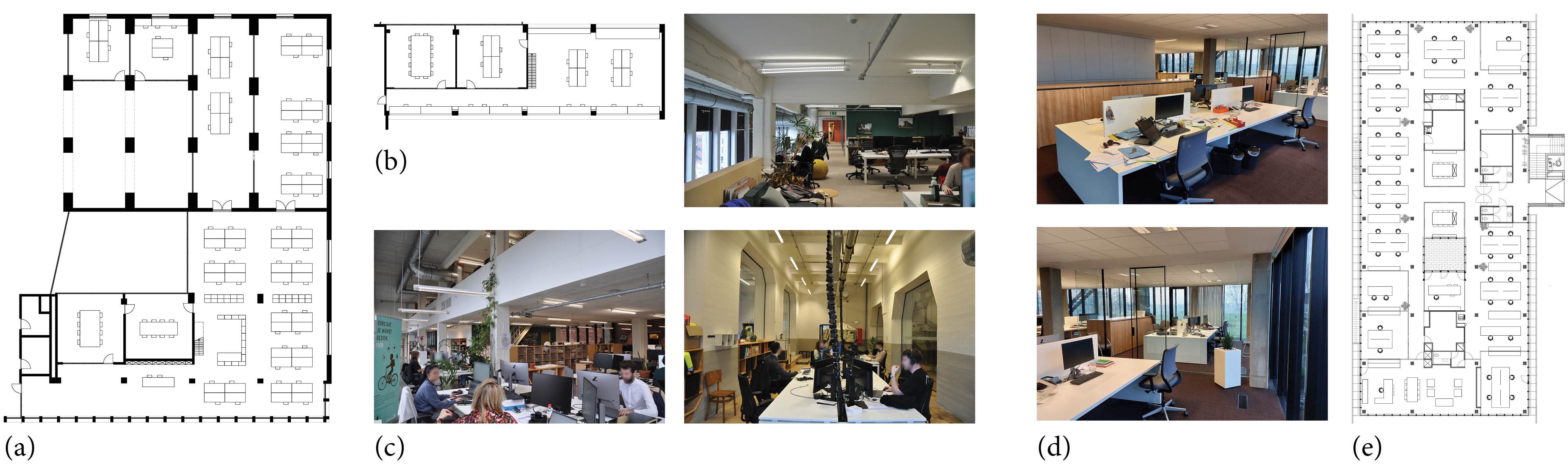}
    \caption{The two studied open-plan office environments: (a) floor plan of office B; (b) mezzanine level of office B; pictures of office B; (d) pictures of office A; and (e) floor plan of office A. Image credits belong to the authors.}
    \Description{ A composite image showing two open-plan office environments. Panel (a) shows the floor plan of office B, and panel (b) presents the mezzanine level of office B, accompanied by pictures of its interior featuring desks, chairs, and open workspaces. Panel (d) contains pictures of office A, showing workstations and office setups, and panel (e) displays the floor plan of office A, detailing desk and room layouts. Both environments are illustrated with architectural plans and real-life photographs for context. All image credits belong to the authors.}
    \label{fig:office-plans}
\end{figure*}

\begin{figure*}[htbp]
    \centering
    \includegraphics[width=0.8\linewidth]{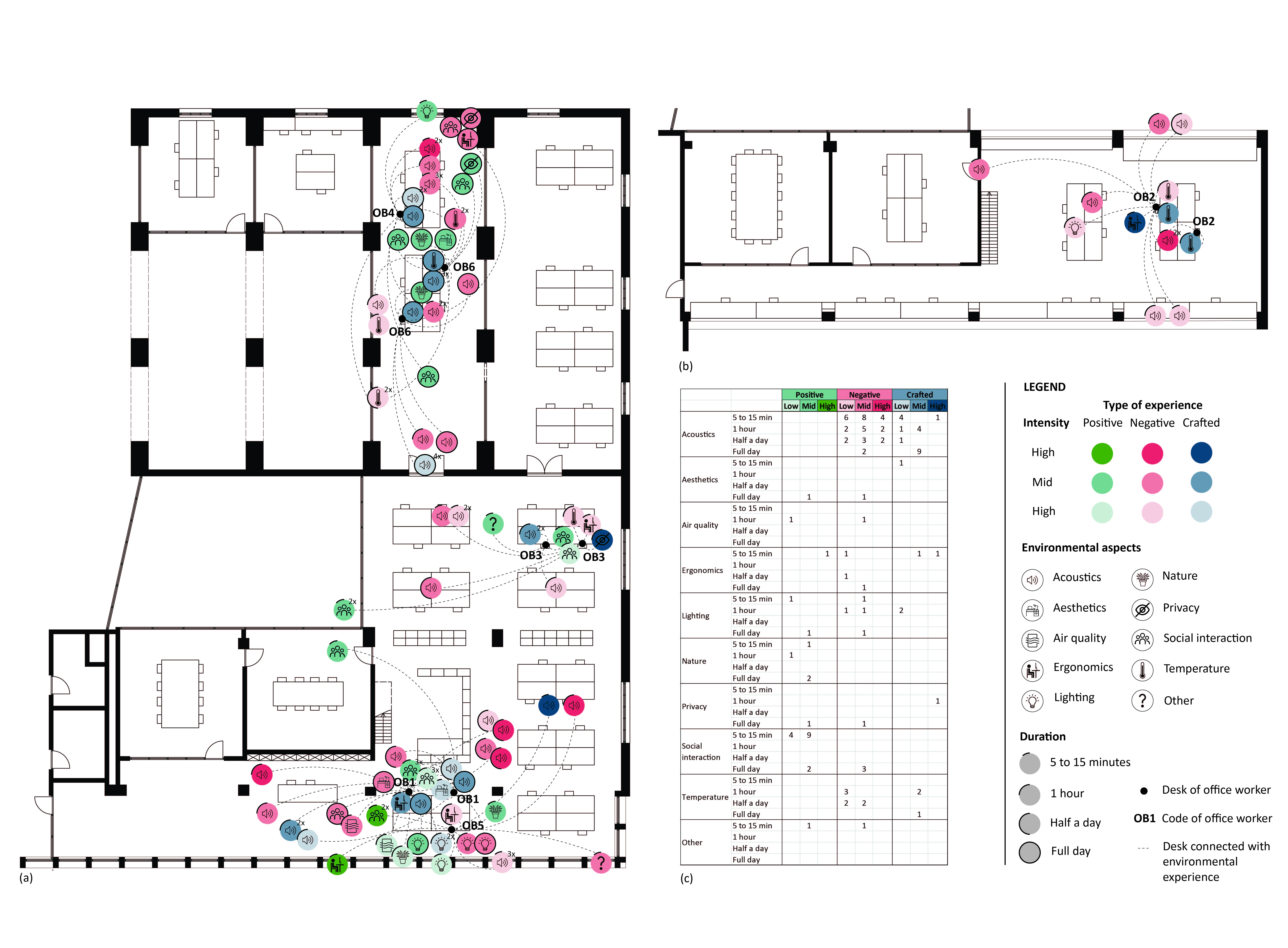}
    \caption{By visually representing the situated experiences mapped by 8 office workers from office B and its mezzanine, we are able to synthesise that: (a) acoustic distractions mainly originate around the desks of other co-workers, positive interactions are experienced at one's own desk, while the presence of nature is experienced in the shared spaces; (b) the individual approaches and frequencies for mapping experience and intensity reveal insights into office workers' personalities; (c) the most frequently mapped environmental aspects are acoustics, lighting and temperature, and they are mapped in relation to their duration and intensity.}
    \Description{A heatmap visualising mapped situated experiences of environmental comfort in office B. Panel (a) shows the heatmap for the main floor of office B, while panel (b) displays the mezzanine level's heatmap. Panel (c) includes a table summarising the number of experiences categorised by environmental aspect, type of experience, intensity and duration. The visuals highlight areas of varying comfort levels within the office spaces.}
    \label{fig:heatmap-appendix}
\end{figure*}



\begin{figure*}[htbp]
    \centering
    \includegraphics[width=1\linewidth]{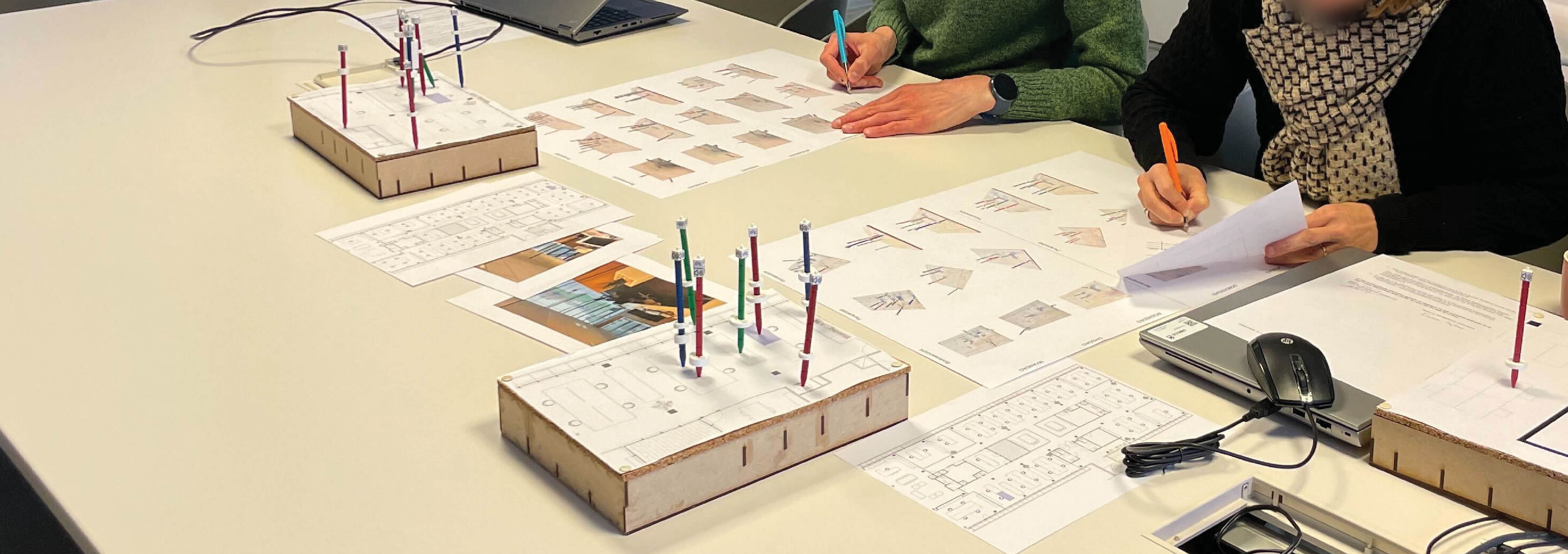}
    \caption{Although in both data interpretation sessions, experts interpreted images of the physical artefacts sent by office workers in relation to office floor plan and images, they also used a physical toolkit to better understand the mapping approaches of the office workers.}
    \Description{An image showing the hands of two people making annotations on sheets of paper with printed images. On the table beside them, there are two boxes. The top lid of both boxes features a printed floor plan, with several pins vertically inserted into it.}
    \label{fig:data-interpretation}
\end{figure*}

\begin{figure*}[htbp]
    \centering
    \includegraphics[width=0.9\linewidth]{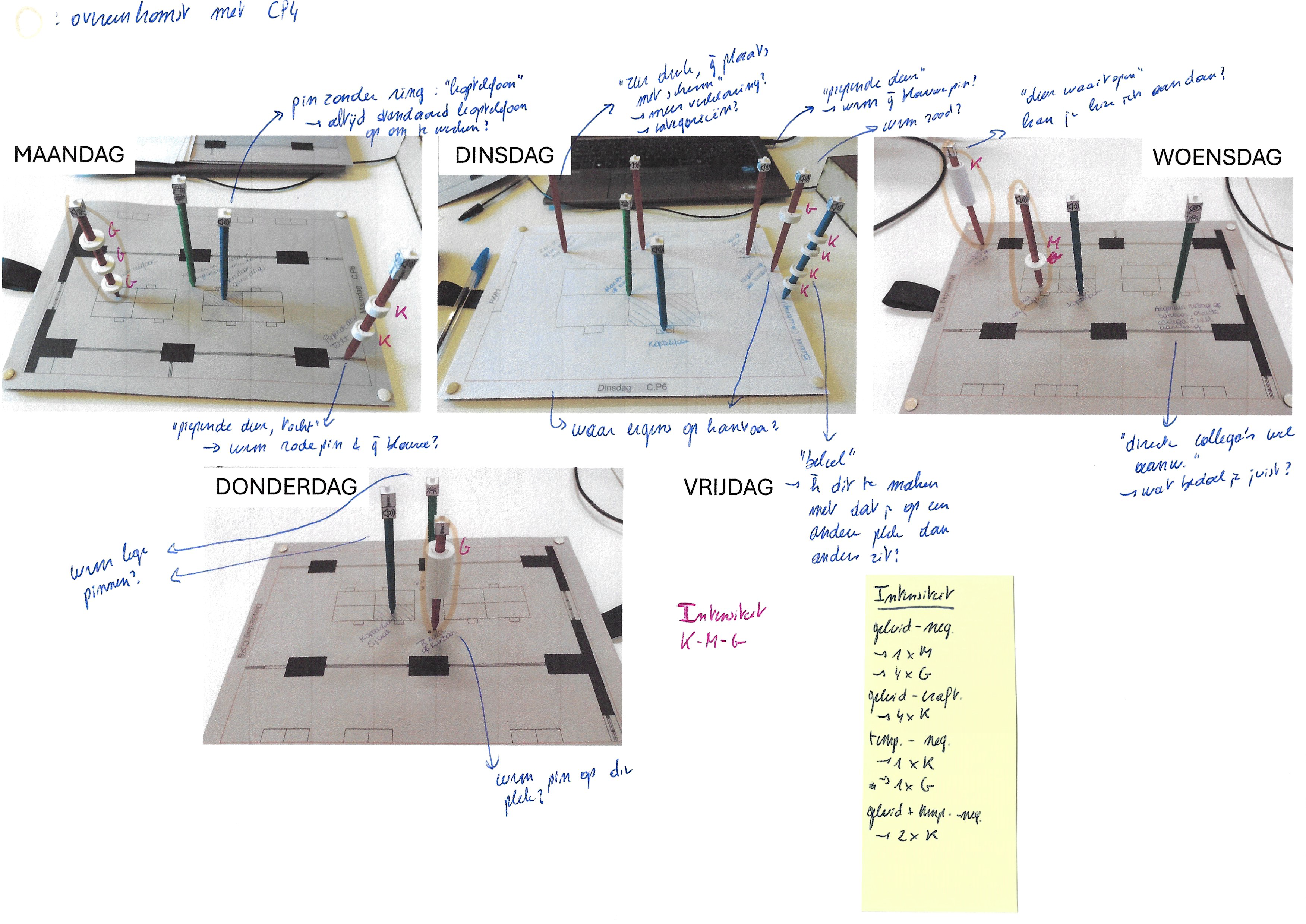}
    \caption{Experts interpreted the toolkit by making annotations next to the artefact images. In this analysis of one office worker from office B, experts question the consistency and interpretation of pin heights, colours, and locations, noting potential ambiguities in representation. Annotations also highlight issues of visibility, awareness, and collaboration, suggesting that different participants may perceive and record the same space in varied ways.}
    \Description{A composite figure consisting of four annotated images, each displaying a different floor plan. Each floor plan has between three and seven pins vertically inserted into its surface. Handwritten annotations with arrows point to specific parts of each image.}
    \label{fig:expert-annotations}
\end{figure*}

\end{document}